\documentclass[12pt]{article}
\usepackage{epsfig,amsfonts,amssymb}
\usepackage{graphicx}
\usepackage{epsfig}
\usepackage{hyperref}
\usepackage{cite}
\def\ZZZ{{\hbox{ Z\kern-1.6mm Z}}}
\def\RRR{{\hbox{ R\kern-2.4mm R}}}
\def\CCC{{\hbox{ C\kern-2.0mm C}}}
\def\zzz{{\hbox{z\kern-1mm z}}}

\newcommand{\qeq}{{\hbox{=\kern-2.3mm ? \kern.5mm }}}
\renewcommand{\qeq}{=}

\newcommand{\eps}{\epsilon}

\newcommand{\wt}{\widetilde}

\newcommand{\be}{\begin{equation}}
\newcommand{\ee}{\end{equation}}
\newcommand{\ben}{\begin{eqnarray}\displaystyle}
\newcommand{\een}{\end{eqnarray}}

\newcommand{\refb}[1]{(\ref{#1})}

\newcommand{\sectiono}[1]{\section{#1}\setcounter{equation}{0}}

\def\one{{\hbox{ 1\kern-.8mm l}}}
\def\zero{{\hbox{ 0\kern-1.5mm 0}}}

\newcommand{\bea}[1]{\begin{eqnarray}\label{#1} }
\newcommand{\eea}{\end{eqnarray}}

\newcommand{\eq}[1]{(\ref{#1})}

\begin{document}

\baselineskip 24pt

\begin{center}
{\Large \bf  S-duality Improved Perturbation Theory in Compactified Type I / Heterotic String Theory}

\end{center}

\vskip .6cm
\medskip

\vspace*{4.0ex}

\baselineskip=18pt

\centerline{\large \rm Roji Pius and Ashoke Sen}

\vspace*{4.0ex}

\centerline{\large \it Harish-Chandra Research Institute}
\centerline{\large \it  Chhatnag Road, Jhusi,
Allahabad 211019, India}

\vspace*{1.0ex}
\centerline{\small E-mail:  rojipius@mri.ernet.in, sen@mri.ernet.in}

\vspace*{5.0ex}

\centerline{\bf Abstract} \bigskip

We study the mass of the stable non-BPS state in type I / heterotic string theory 
compactified on
a circle with the help of the interpolation formula between weak and strong coupling
results. 
Comparison between the results at different orders indicate that this procedure can
determine the mass of the particle to within 10\% accuracy over the entire two dimensional
moduli space parametrized by the string coupling and the radius of compactification.
This allows us to estimate the region of the stability of the particle in this two
dimensional moduli
space. Outside this
region the particle is unstable against decay into three BPS states carrying the same total
charge as the original state. We  discuss generalization of this analysis to 
compactification on higher
dimensional tori.

\vfill \eject

\baselineskip=18pt

\tableofcontents

\sectiono{Introduction} \label{sint}

Our current understanding of string theory is based mostly on perturbation expansion
in the string coupling\cite{1209.5461}. 
Furthermore this perturbation expansion is believed to be an asymptotic
expansion. For this reason one might worry that our ability to compute anything in string theory
may be limited to very narrow corners of the full string theory landscape -- regions in which the
theory admits a description as a very weakly coupled string/M/F-theory.

Ref.~\cite{1304.0458} suggested making use of duality and suitable interpolation formula
to translate the weak coupling results in string theory to approximate results for physical quantities
over the entire range of string coupling constant. As a specific example, the mass of the
stable non-BPS particle in ten dimensional  type I / SO(32) heterotic string theory was considered. Using
a suitable formula that interpolates between the result for this mass in weakly coupled
SO(32) heterotic string theory and weakly coupled type I string theory, an approximate formula
for the mass of this state was derived over the entire range of string coupling. Furthermore by
comparing the results in different orders it was estimated that this approximate formula lies within
10\% of the exact result over the entire range of coupling. Generalization of this analysis has since
been discussed in \cite{1306.3228,1307.3689,1310.3757}.

Since  a generic string theory moduli space is multi-dimensional it is natural to ask if this interpolation
technique can be used to find approximate expressions for various physical quantities over the
full multi-dimensional moduli space.\footnote{This issue arose aleady in the analysis of \cite{1306.3228}
({\it albeit} in the context of a supersymmetric gauge theory instead of string theory) where the
interpolation technique together with perturbative results were used to determine approximate
formul\ae\ for anomalous dimension of non-BPS operators in the complex coupling constant plane.}
In this paper we explore this in the context of SO(32) heterotic / type I string theory compactified
on a circle. If we do not switch on any Wilson line so that the SO(32) gauge group is unbroken then
the moduli space is two dimensional, parametrized by the string coupling and the radius of 
compactification. We use perturbative results in the compactified string theory and a suitable
generalization of the interpolation formula used in \cite{1304.0458} to derive expression for the
mass of the non-BPS state in the full two dimensional moduli space. Comparison between different
orders of approximation again indicates that the approximate formula derived here lies within 10\%
of the exact formula over the entire two dimensional moduli space. 

In type I / SO(32) heterotic string theory compactified on $S^1$, one can identify a set of BPS
states whose total charge is the same as that of the charge carried by the non-BPS state under
study. Thus the latter can decay into the former if the mass of the non-BPS state is larger than
the sum of the masses of the BPS states to which it could possibly decay. With the help of the
approximate formula for the mass we determine the part of the region of the two dimensional 
moduli space in which the non-BPS state is unstable. Again we find that the region determined this
way is only mildly sensitve to the order of the approximation that we use. 

The rest of the paper is organized as follows. In \S\ref{snorm} we review the interpolation
technique and the normalization conventions of \cite{1304.0458}, and the additional ingredients
we need for dealing with the compactified theory. In \S\ref{stypeI} we carry out the computation of
the one loop correction to the mass of stable non-BPS state in type I string theory compactified
on $S^1$. In \S\ref{sheterotic} we carry out a similar calculation in SO(32) heterotic string
theory compactified on $S^1$. In \S\ref{sinterpol} we construct the interpolation formula for the
mass of the non-BPS state in various approximation, and compare the results in different orders
of approximation. In \S\ref{sstability} we use the interpolation formula to analyze the region
of stability of the non-BPS state. In \S\ref{shigher} we discuss extension of our analysis to
the case of compactification on higher dimensional tori. We conclude in \S\ref{sconc} with a
brief summary of our results and their possible relation to related developments.

\sectiono{Normalization conventions and tree level results} \label{snorm}

We shall use the normalization conventions used in \cite{1304.0458 }. 
The purpose of this section will be to review these conventions and also introduce
the extra conventions involving the radius of compactification.

Let $g_H$ and $g_I$  be the string coupling in ten dimensional
 heterotic and type I  string theories respectively. 
 We introduce a new coupling parameter $g$ in terms of which $g_H$ and $g_I$
 are given by
 \be \label{edefg}
g_H = 2^{7/2} \pi^{7/2} g, \quad g_I = 2^{3/2} \pi^{7/2} g^{-1}\, .
\ee
We normalize the heterotic and type I metric such that 
the heterotic string tension in heterotic metric
and the type I string tension in type I metric are both given by $1/2\pi$.
The mass of the non-BPS state, measured in the ten dimensional 
Einstein metric, is parametrized by
\be \label{edefF}
M(g) = 2^{15/8} \pi^{7/8} F(g)\, .
\ee
The tree level weak and strong coupling values of $F(g)$, computed respectively
from tree level heterotic and type I string theories, are
\be  \label{eleading}
F^W_0(g) = g^{1/4}, \qquad F^S_0(g) = g^{3/4}\, .
\ee

Upon compactification on a circle the tree level masses will continue to be given by
\refb{eleading} if we measure it in the canonical metric in ten dimensions. We shall follow
this convention. 
Let $r_I$, $r_H$ and $r_E$ denote the radii of the compact circle measured in the type I, heterotic
and canonical metric. Then
\ben \label{erirh}
r_E &=& (g_I)^{-1/4} r_I = 2^{-3/8} \pi^{-7/8} g^{1/4} r_I\, , \nonumber \\
r_H &=& (g_H)^{1/4} r_E = 2^{7/8} \pi^{7/8} g^{1/4} r_E = 2^{1/2} g^{1/2} r_I\, .
\een

We expect the quantum corrections in the heterotic and type I string theories to modify 
the weak coupling results to
\be \label{expan}
F^W(g) = g^{1/4} \left[1+ \sum_{k=1}^\infty\, A_{2k}(r_H) g^{2k}\right], \qquad 
F^S(g) = g^{3/4} \left[1+ \sum_{n=1}^\infty\, B_n(r_I) g^{-n}\right], 
\ee
where the functions $A_{2k}(r_H)$ and $B_n(r_I)$ have finite $r_H\to\infty$ and 
$r_I\to\infty$ limits respectively, corresponding to the results in the non-compact theory.
We now introduce the interpolating function\footnote{There are various other possible
interpolation schemes (see {\it e.g,} \cite{0706.1555,kleinert}), 
but the one given in \refb{einterpol}, called the fractional power
of polynomial (FPP) scheme in \cite{1306.3228}, 
seems to be most suitable for our purpose as this gives a 
clear separation between the coefficients which are determined using weak coupling expansion and
the coefficients which are determined using strong coupling expansion.
This is needed to ensure that the weak coupling expansions {\it at fixed $r_H$} and strong
coupling expansion {\it at fixed $r_I$} match the perturbation expansions. The difficulty in
achieving this with other approximation schemes, {\it e.g.} 2-point Pad\'e approximant, is
similar to the difficulties faced 
in \cite{1306.3228} in getting a duality invariant approximation scheme beyond
four loops using 2-point Pad\'e approximant.}
\ben \label{einterpol}
F_{m,n}(g) &=& g^{1/4} \bigg[1 + a_1(r_H) g +\cdots a_m(r_H) g^{m} +
b_{n}(r_I) g^{m+1} + b_{n-1}(r_I) g^{m+2} \nonumber \\ && \qquad
+ \cdots + b_1(r_I) g^{m+n} 
+  
g^{m+n+1}\bigg]^{1/\{2(m+n+1)\}}\, ,
\een
where the functions $a_k(r_H)$ and $b_k(r_I)$ are determined as follows.
We determine $a_k(r_H)$ by demanding that 
after setting the $b_k$'s to zero,
the expansion of
\refb{einterpol} in powers of $g$ at fixed $r_H$
agrees with that of $F^W(g)$ up to order $g^{{1\over 4}+m}$. Similarly
the functions $b_k(r_I)$ are determined by demanding that
after setting the $a_k(r_H)$'s to zero, 
the expansion of
\refb{einterpol} in powers of $g^{-1}$ at fixed $r_I$ 
agrees with that of $F^S(g)$ up to order $g^{{3\over 4}-n}$.

We shall now argue that the weak coupling expansion at fixed $r_H$ 
of the full function $F_{m,n}$
keeping both $a_k$'s and $b_k$'s non-zero coincides with that of $F^W$
up to order $g^{{1\over 4}+m}$, and similary the strong coupling expansion at fixed $r_I$
of the
full function $F_{m,n}$  coincides with that of $F^S$
up to order $g^{{3\over 4}-n}$. From eq.\refb{erirh} it follows that 
 $r_I = 
2^{-1/2} g^{-1/2} r_H$, and hence as $g\to 0$ keeping $r_H$ fixed, $r_I\to\infty$.
In this limit the coefficients $B_k$ appearing in the strong coupling expsnsion should
approach finite values given by the results in the non-compact theory.
Thus the coefficients $b_k$, determined in terms of the coefficients $B_\ell$ for
$k\le n$, should also approach finite values in this limit. This shows that the expansion
of $b_k$ in powers of $g$ at fixed $r_H$ contains non-negative powers of $g$.
Substituting this into \refb{einterpol} we now see that the coefficents $b_k$ do not affect
the weak coupling expansion of $F_{m,n}$ to order $g^{{1\over 4}+m}$, and hence 
the weak coupling expansion of $F_{m,n}$ to this order
agrees with that of $F^W$. A similar analysis shows that the expansion of $a_k$ in powers
of $g^{-1}$ at fixed $r_I$ contains non-positive powers of $g$. Hence the 
expansion of $F_{m,n}$ in powers of $g^{-1}$ at fixed $r_I$ 
to order $g^{{3\over 4}-n}$ is insensitive to the coefficients $a_k$ and coincides
with that of $F^S(g)$.

{}From \refb{eleading}, \refb{einterpol} 
we can find the following interpolating functions for the mass of the
non-BPS particle
\ben
F_{0,0}(g) &=& g^{1/4} \, (1 + g)^{1/2}\, , \nonumber \\
F_{1,0}(g) &=& g^{1/4} \, (1 + g^2)^{1/4}\, .
\een

\sectiono{Strong coupling expansion } \label{stypeI}

 Denote by $\Delta M$ the first order correction to the mass formula from the strong coupling 
 end, i.e. in 
 weakly coupled 
 type I string theory compactified on a circle $S^1$ of radius $r_I$. This can be obtained by calculating
 the one loop correction to the energy of the non-BPS D0-brane of type I string theory compactified on $S^1$. 
This calculation differs from the corresponding calculation in \cite{1304.0458} by having to include
extra contribution from open string winding modes along the circle, begining on a D0-brane and ending on
one of its images. 
The result takes the form\cite{9510017,9903123,0003022,0012167}
 \ben\label{openchannel}
 -\Delta M = \frac{1}{2}g_I^{\frac{1}{4}}(8\pi^2)^{-\frac{1}{2}}\int_0^{\infty} s^{-\frac{3}{2}}ds[Z_{NS;D0D0}-Z_{R;D0D0}+Z_{NS;D0D9}-Z_{R;D0D9}],
 \een
where $Z_{NS;D0D0},Z_{R;D0D0},Z_{NS;D0D9},Z_{R;D0D9}$ denote respectievely the contributions from the NS and R sector open strings with both ends on the D0-brane and NS and R open strings with one end on the D0-brane
and the other end on the D9-brane wrapped on $S^1$ of radius $r_I$. Explicit computation gives
\ben
Z_{NS;D0D0} &=& \frac{1}{2}\Bigg(\sum_n \tilde{q}^{2 n^2r_I^2}\Bigg)\frac{f_3(\tilde{q})^8}{f_1(\tilde{q})^8}+2^{\frac{5}{2}}(1-i)\frac{f_3(i\tilde{q})^9f_1(i\tilde{q})}{f_2(i\tilde{q})^9f_4(i\tilde{q})}-2^{\frac{5}{2}}(1+i)\frac{f_4(i\tilde{q})^9f_1(i\tilde{q})}{f_2(i\tilde{q})^9f_3(i\tilde{q})},\cr
Z_{R;D0D0} &=& \frac{1}{2}\Bigg(\sum_n \tilde{q}^{2 n^2r_I^2}\Bigg)\frac{f_2(\tilde{q})^8}{f_1(\tilde{q})^8},\cr
Z_{NS;D0D9} &=& 16\sqrt{2}\frac{f_2(\tilde{q})^9f_1(\tilde{q})}{f_4(\tilde{q})^9f_3(\tilde{q})},\cr
Z_{R;D0D9} &=& 16\sqrt{2}\frac{f_3(\tilde{q})^9f_1(\tilde{q})}{f_4(\tilde{q})^9f_2(\tilde{q})},
\een
where $n$ is the quantum number corresponding to the winding number of the fundamental open
string
along the compact direction and
\ben
\tilde{q} \equiv e^{-\pi s},
\een
\ben
f_1(q) &=& q^{1/12}\prod_{n=1}^{\infty}(1-q^{2n}) = \eta(2\tau),\quad q \equiv e^{2\pi i \tau},\cr
f_2(q) &=& \sqrt{2} \, q^{1/12}\prod_{n=1}^{\infty}(1+q^{2n}) = \sqrt{2}\, \frac{\eta(4\tau)}{\eta(2\tau)},\cr
f_3(q) &=& q^{-1/24}\prod_{n=1}^{\infty}(1+q^{2n-1}) = \frac{\eta(2\tau)^2}{\eta(\tau)\eta(4\tau)},\cr
f_4(q) &=& q^{-1/24}\prod_{n=1}^{\infty}(1-q^{2n-1}) = \frac{\eta(\tau)}{\eta(2\tau)}.
\een

Individual terms in \eq{openchannel} are both IR and UV divergent. Using the prescription for the IR and UV regularization described in \cite{1304.0458}
we can express \eq{openchannel} as follows,
\be
\Delta M = \tilde K_S (g_I)^{1/4}  \, ,
\ee
\ben\label{open}
\tilde K_S &\equiv& -\frac{1}{2}(8\pi^2)^{-\frac{1}{2}} \lim_{\Lambda \to \infty} \lim_{\epsilon\to 0} \Bigg[\int_{\eps}^{\Lambda}s^{-\frac{3}{2}}ds\Bigg\{\frac{1}{2}\Bigg(\sum_n \tilde{q}^{2n^2r_I^2}\Bigg)\Bigg(\frac{f_3(\tilde{q})^8}{f_1(\tilde{q})^8}-\frac{f_2(\tilde{q})^8}{f_1(\tilde{q})^8}\Bigg) 
\cr 
 && +16\sqrt{2}\frac{f_2(\tilde{q})^9f_1(\tilde{q})}{f_4(\tilde{q})^9f_3(\tilde{q})}-16\sqrt{2}\frac{f_3(\tilde{q})^9f_1(\tilde{q})}{f_4(\tilde{q})^9f_2(\tilde{q})}\Bigg\} 
 \cr 
 && 
 +\int_{\eps/4}^{\Lambda}s^{-\frac{3}{2}}ds\Bigg\{ 2^{\frac{5}{2}}(1-i)\frac{f_3(i\tilde{q})^9f_1(i\tilde{q})}{f_2(i\tilde{q})^9f_4(i\tilde{q})} -2^{\frac{5}{2}}(1+i)\frac{f_4(i\tilde{q})^9f_1(i\tilde{q})}{f_2(i\tilde{q})^9f_3(i\tilde{q})}\Bigg\}\Bigg]\, .
\een
Note that for $r_I < 1/\sqrt 2$, the $n=1$ term in the sum behaves as $\wt q^{2 r_I^2 -1}
= e^{\pi s (1 - 2 r_I^2)}$ and hence the integral over $s$ has a divergence from the large $s$
region. This reflects the appearance of the open string tachyon in the spectrum for 
$r_I<1/\sqrt 2$\cite{9904207}.
For this reason the open string loop corrections to the mass of stable non-BPS state makes sense
only for $r_I\ge 1/\sqrt 2$, and in the rest of this section we shall focus on this region. Using 
\refb{erirh} we see that in terms of the
radius $r_H$ in the heterotic metric, this condition takes the form
\be \label{erhcond1}
r_H > g^{1/2}\, .
\ee

It is possible to convert expression \eq{open} in the \textquoteleft closed string channel\textquoteright\  
using the modular tranformation laws of $f_i$'s :
\ben\label{closedchannel}
\tilde{K_S} = - \lim_{\Lambda \to \infty} \lim_{\epsilon\to 0}\frac{1}{4\pi}(8\pi^2)^{-\frac{1}{2}}\Bigg[\int_{\pi/\Lambda}^{\pi/\eps}
dt(C_{00}+C_{09}+C_{09}^{*})+\int_{\pi/4\Lambda}^{\pi/\eps}dt(\cal{M}+\cal{M^{*}})\Bigg]
\een
where
\ben
C_{00}&=& \Bigg(\frac{\pi}{t}\Bigg)^4\Bigg(\sum_n\frac{q^{\frac{n^2}{2 r_I^2} }}{\sqrt 2 \, r_I}\Bigg)\Bigg(\frac{f_3(q)^8}{f_1(q)^8}-\frac{f_4(q)^8}{f_1(q)^8}\Bigg), \cr
C_{09} &=& 2^{\frac{9}{2}}\Bigg(\frac{f_4(q)^9f_1(q)}{f_2(q)^9f_3(q)} - \frac{f_3(q)^9f_1(q)}{f_2(q)^9f_4(q)} \Bigg), \cr
\cal{M} &=& 2^{\frac{9}{2}}\Bigg(\frac{f_3(iq)^9f_1(iq)}{f_2(iq)^9f_4(iq)} - \frac{f_4(iq)^9f_1(iq)}{f_2(iq)^9f_3(iq)}\Bigg),\cr
q &\equiv& e^{-t}.
\een
$C_{00}$ denotes the cylinder amplitude with boundaries lying on the D0-brane, given by the inner product between the boundary states of D0-brane. $C_{09}$ denotes the cylinder amplitude with one boundary lying on the D0-brane and the other boundary on the D9-brane wrapped on $S^1$,
given by the inner product between the boundary states of D0-brane and the D9-brane wrapped on $S^1$. $\cal{M}$ denotes the m\"{o}bius strip amplitude with boundary lying on the D0-brane, given by the inner product between the boundary states of D0-brane and the crosscap. 

Using this and eqs.\refb{edefg}, \refb{edefF}
we can write the corrected strong coupling expression for $F^S(g,r_I)$ to order $g^{{3\over 4}-1}$ as,
\ben \label{efs1g}
F^S_1(g,r_I) = g^{\frac{3}{4}}\bigg(1+K_S(r_I)g^{-1}\bigg),     \quad    K_S(r_I) \equiv 2^{-\frac{3}{2}}\tilde{K_S} \, .
\een
$K_S(r_I)$ can be obtained by  integrating expression \eq{open} numerically for 
different values of $r_I$.  We find that the result of this numerical evaluation fits well with the
function
\be
K_{S}(r_I) \simeq 0.351
-0.048\,  \exp\left[-10 \, (r_I- 2^{-1/2})^{2/3}\right]\, ,
\ee
within 1\% accuracy over the entire range $1/\sqrt{2}\le r_I<\infty$.


\sectiono{Weak coupling expansion} \label{sheterotic}

First order correction to $M$ in the weakly coupled heterotic string theory given by
\ben
\delta M = MK_W(r_H)g^2,
\een
where $K_W(r_H)$ can be calculated by doing a one loop heterotic string calculation similar to
that in \cite{1304.0458}, but including the effect of closed heterotic string winding and momentum
modes along the circle. The result is
\ben\label{weak1}
K_W(r_H)=&&-\frac{1}{64\pi}\int d^2\tau \int d^2z \Bigg[\Bigg\{\sum_{\nu'}\{\overline{\vartheta_{\nu'}(\frac{z}{2})^{16}} \}(\overline{\eta(\tau)})^{-18}(\eta(\tau))^{-6}\Bigg(\frac{\vartheta_{11}(z)}{\overline{\vartheta_{11}(z)}}\Bigg)^2\Bigg\}\cr
&&\textrm{exp}\left(-\frac{4\pi z_2^2}{\tau_2}\right)
(\tau_2)^{-9/2}\frac{1}{r_H}\Bigg\{\sum_{n,w}\textrm{exp}\bigg(-\frac{\pi i \overline{\tau}}{2}(\frac{n}{r_H}+wr_H)^2+\frac{\pi i \tau}{2}(\frac{n}{r_H}-wr_H)^2\bigg)\Bigg\}\Bigg],
\nonumber \\
\een
with $\tau$ denoting the modular parameter of the torus, $\nu$ denoting the spin structure on the torus taking values 00, 01, 10 and 11, $\vartheta$ are the Jacobi theta functions, $r_H$ radius of $S^1$ on which heterotic string theory is compactified and $n,w$
representing the momentum and winding number along the compactified direction. Since this
expression is invariant under T-duality transformation $r_H\to 1/r_H$ (except for the overall factor
of $1/r_H$ that is taken care of by the transformation law of $g^2$ multiplying it), we can focus on
the region $r_H\ge 1$. In this region the evaluation of the integral can be facilitated using a
Poisson resummation in the variable $n$. This yields
\ben\label{weak2}
K_W(r_H)=&&-\frac{1}{64\pi}\int d^2\tau \int d^2z \Bigg[\Bigg\{\sum_{\nu'}\{\overline{\vartheta_{\nu'}(\frac{z}{2})^{16}} \}(\overline{\eta(\tau)})^{-18}(\eta(\tau))^{-6}\Bigg(\frac{\vartheta_{11}(z)}{\overline{\vartheta_{11}(z)}}\Bigg)^2\Bigg\}\cr
&&\textrm{exp}\left(-\frac{4\pi z_2^2}{\tau_2}\right)(\tau_2)^{-5}\Bigg\{\sum_{k,w}\textrm{exp}\bigg(-\frac{\pi}{\tau_2}r_H^2 |k-w\tau|^2 
\bigg)\Bigg\}\Bigg]\, .
\een
In the $r_H\to\infty$ limit only $k=w=0$ term in the sum survives, giving back the ten dimensional
result. For finite $r_H$ numerically
integrating expression \eq{weak2} for different values of $r_H$ we find that the result can be
fitted approximately with the function,
\be 
K_{W}(r_H) \simeq 0.23\, \bigg(1+\frac{1}{r_H^7}\bigg)^{2/7}.
\ee
Then the corrected weak coupling expression for $F^W(g,r_H)$ to order $g^{2}$ is given as,
\ben \label{efw2grh}
F^W_2(g,r_H) = g^{\frac{1}{4}}\bigg(1+K_W(r_H)g^{2}\bigg)\, .
\een


Notice that $K_{W}(r_H)$ diverges in the $r_H\to 0$ limit. This is easily understood using the
known T-duality invariance $r_H\to 1/r_H$ in the heterotic string theory. Under this the ten
dimensional string coupling transforms to $g/r_H$. Defining
\be \label{etdual}
\wt r_H = {1\over r_H}, \quad \wt g = {g\over r_H}\, ,
\ee
we can express \refb{efw2grh} as
\be \label{efw3}
F^W_2(g,r_H) = (\wt r_H)^{-1/4}\, 
\wt g^{\frac{1}{4}}\bigg(1+.23\, (1 + (\wt r_H)^{-7})^{2/7} \wt g^{2}\bigg)\, .
\ee
Except for the overall factor of $(\wt r_H)^{-1/4}$ which reflects the overall scale factor relating the
ten dimensional Einstein metric in the dual pair of heterotic string theories, we see that this has a 
perfectly good $\wt r_H\to \infty$ ($r_H\to 0$) limit at fixed $\wt g$.  For this reason, for $r_H<1$
it is more natural to use the coupling constant $\wt g$ of the T-dual theory as an expansion parameter.

\sectiono{Interpolating functions} \label{sinterpol}

\begin{figure}
\begin{center}
\epsfysize=5cm
\epsfbox{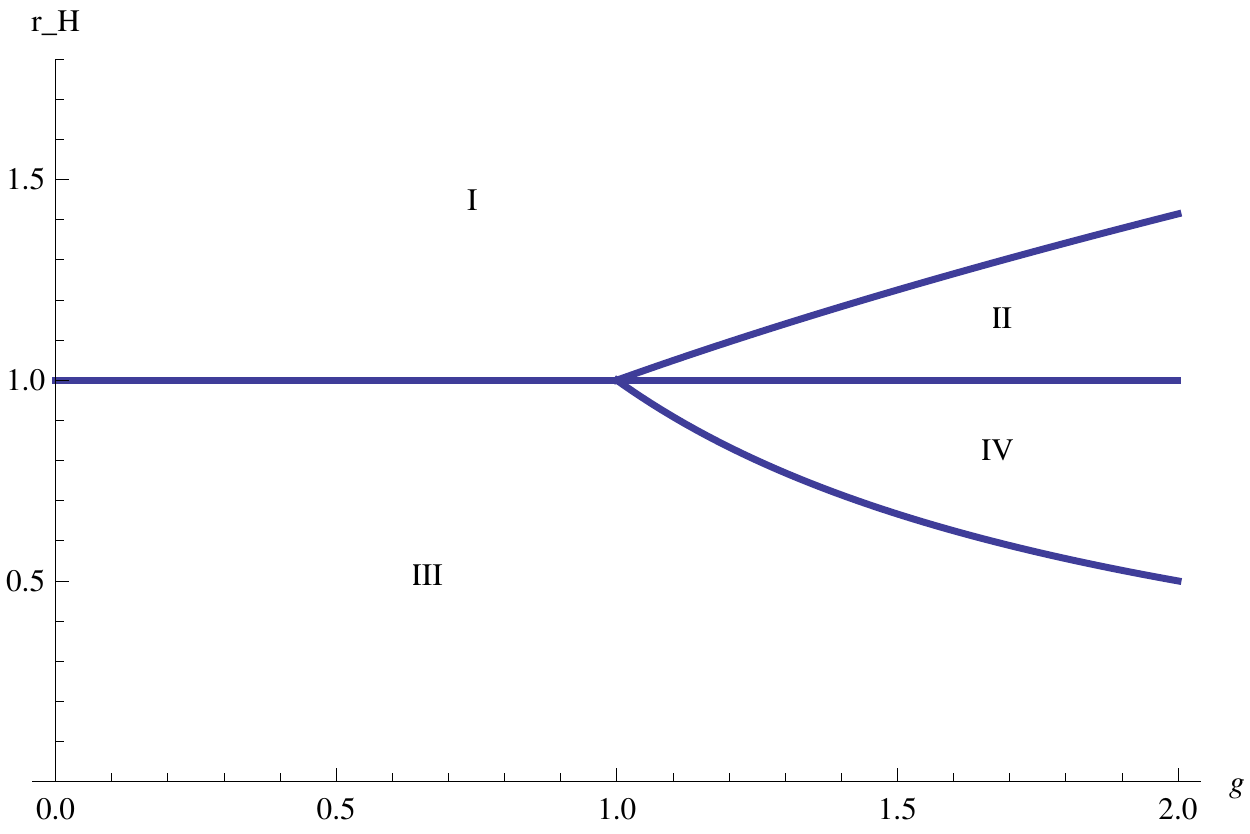}
\end{center}
\caption{The four regions in the $g$-$r_H$ plane. The curves bounding these regions are
$r_H=1$, $r_H=g^{1/2}$ and $r_H=g^{-1}$. \label{f3}}
\end{figure}

We now turn to the construction of the interpolating functions. For definiteness we shall treat $g$ and
$r_H$ as independent variables. We shall divide up the $g$-$r_H$ plane into several regions shown
in Fig.~\ref{f3} and use
different interpolating functions in these different regions. 

\medskip

\noindent{\bf Region I:} First consider the region I defined by
\ben
{\bf I} &:&  r_H\ge 1, \quad r_H\ge g^{1/2}\, .
\een
In this region $r_I\ge 1/\sqrt 2$ (see \refb{erhcond1}) and
both the heterotic perturation theory in powers of $g$ 
and type I perturbation theory in powers of $g^{-1}$ are well defined at small and
large $g$ respectively.
Thus we can use standard interpolation formula described in \S\ref{snorm}:
\be
F_{0,0}(g,r_H) = g^{1/4} \, (1 + g)^{1/2}\, , 
\ee
\be
F_{1,0}(g,r_H) = g^{1/4} \, (1 + g^2)^{1/4}\, ,
\ee
\be
F_{0,1}(g,r_H)= g^{1/4}\, (1 + 4\,  K_S(r_I) g +g^2)^{1/4}\, ,
\ee
\be 
F_{1,1}(g,r_H) =g^{1/4}\, (1 + 6\,  K_S(r_I) g^2 +g^3)^{1/6}\, ,
\ee
\be 
F_{2,0}(g,r_H) =  g^{1/4}\, (1 + 6\, K_W(r_H) g^2 +g^3)^{1/6}\, ,
\ee
\be 
F_{2,1}(g,r_H) =  g^{1/4}\, (1 + 8\, K_W(r_H) g^2 + 8 K_S(r_I) g^3 +g^4)^{1/8}\, ,
\ee
\be 
F_{3,0}(g,r_H) =  g^{1/4}\, (1 + 8\, K_W(r_H) g^2 +g^4)^{1/8}\, ,
\ee
and 
\be
F_{3,1}(g,r_H) = g^{1/4}\, \left(1 + 10\, K_W(r_H) g^2 +10 \, K_S(r_I)  g^4 +g^5\right)^{1/10}\, .
\ee

\medskip

\noindent{\bf Region II:} Region II is defined by
\ben
{\bf II} &:&  1\le r_H\le g^{1/2}\, .
\een
In this region heterotic perturbation theory in power of $g$ is still valid for small $g$
but type I perturbation theory in powers of $g^{-1}$ breaks down at large $g$
due to the presence of the tachyon. Thus the only interpolating functions we can use are
$F_{m,0}$ for $0\le m\le 3$.

\medskip

\noindent{\bf Region III:} Region III is defined by
\ben \label{er3}
{\bf III} &:&  r_H\le 1, \quad r_H\le g^{-1}\, .
\een
The significance of this region can be understood by reexpressing \refb{er3} in
terms of T-dual variables $\wt r_H$, $\wt g$ introduced in \refb{etdual}.
This corresponds to
\be 
\wt r_H\ge 1, \quad \wt r_H \ge \wt g^{1/2}\, .
\ee
In this region we shall use the interpolating functons of region I with $(g, r_H)$ replaced by
$(\wt g, \wt r_H)$ and with an overall multiplicative factor of $(\wt r_H)^{-1/4}= (r_H)^{1/4}$ to account for
the rescaling of the canonical metric discussed below \refb{efw3}. Thus we use the functions
\be
\wt F_{m,n}(g, r_H) = (r_H)^{1/4} \, F_{m,n}(g/r_H, 1/r_H)\, .
\ee
Physically this corresponds to using an interpolating formula between the T-dual heterotic string
theory and its strong coupling dual type I string theory (obtained in the $\wt g\to\infty$ limit at
fixed $\wt r_H /\wt g^{1/2}$). This is not the original type I string theory, but related to it via a
strong-weak coupling duality transformation. This is apparent from the fact that while in the original
type I string theory the non-BPS D-brane develops a tachyon for $r_H < g^{1/2}$, in the new theory
the non-BPS D-brane is tachyon free for $r_H < g^{-1}$. 

\medskip

\noindent{\bf Region IV:} Region IV is defined by
\ben \label{er35}
{\bf IV} &:&  g^{-1} \le r_H\le 1\, .
\een
In the $(\wt g, \wt r_H)$ variables this corresponds to $1\le \wt r_H\le \wt g^{-1/2}$, \i.e.\ this is
the heterotic T-dual image of region II. Thus we use the interpolation formul\ae\ of region II with
$(g,r_H)$ replaced by $(\wt g, \wt r_H)$:
\be
\wt F_{m,0}(g, r_H) = (r_H)^{1/4} \, F_{m,0}(g/r_H, 1/r_H)\, .
\ee

Note that the results in regions III and IV can be obtained from those in regions I and II by
heterotic T-duality transformation \refb{etdual}. For this reason we shall focus on regions I and II from
now on.

\begin{figure}
\vskip -1in
\begin{center}
\epsfysize=5cm
\hbox{\epsfbox{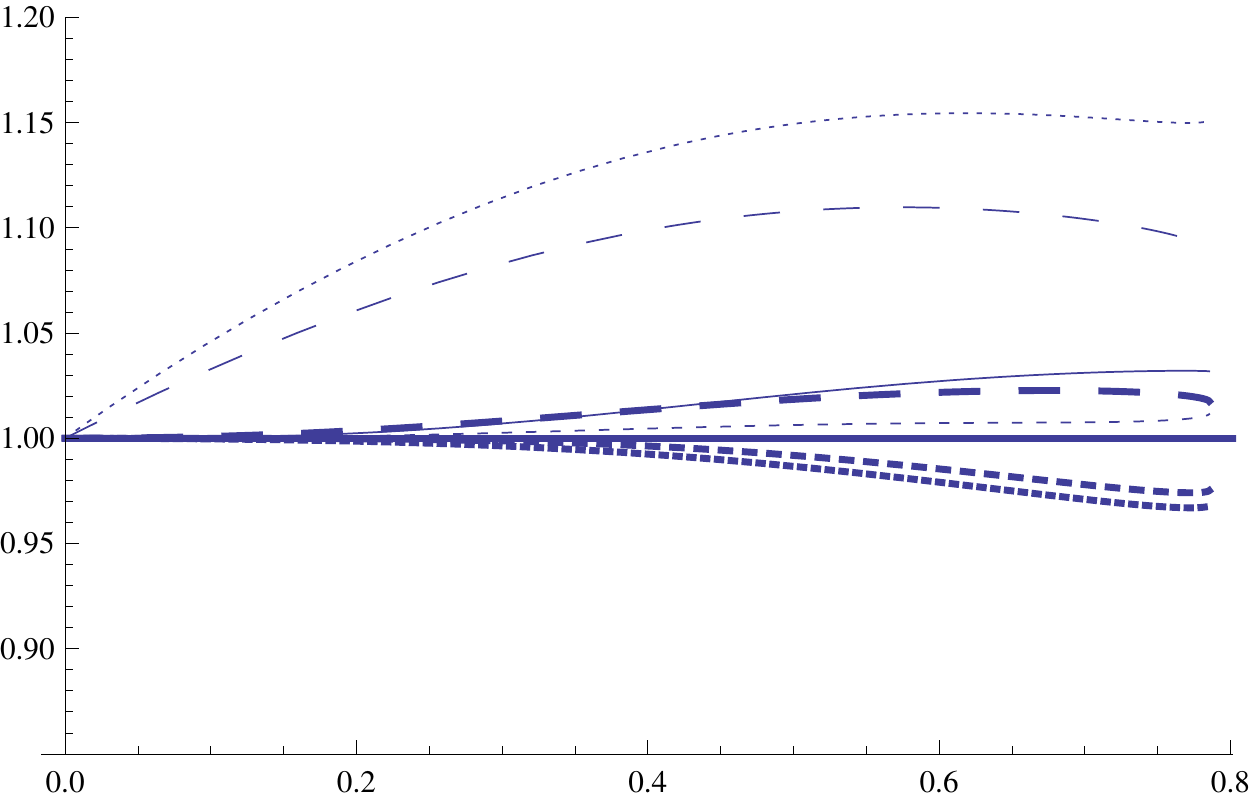} \epsfysize=5cm \epsfbox{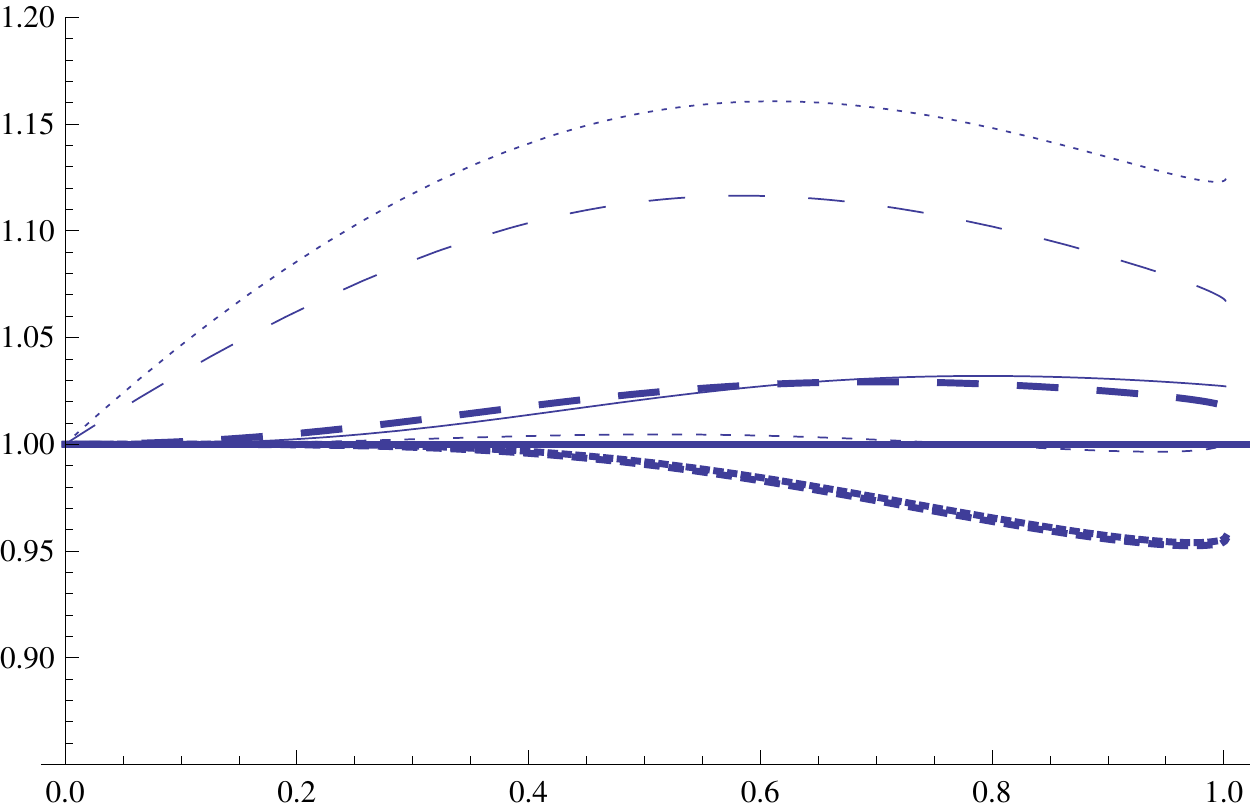}}
\epsfysize=5cm
\hbox{\epsfbox{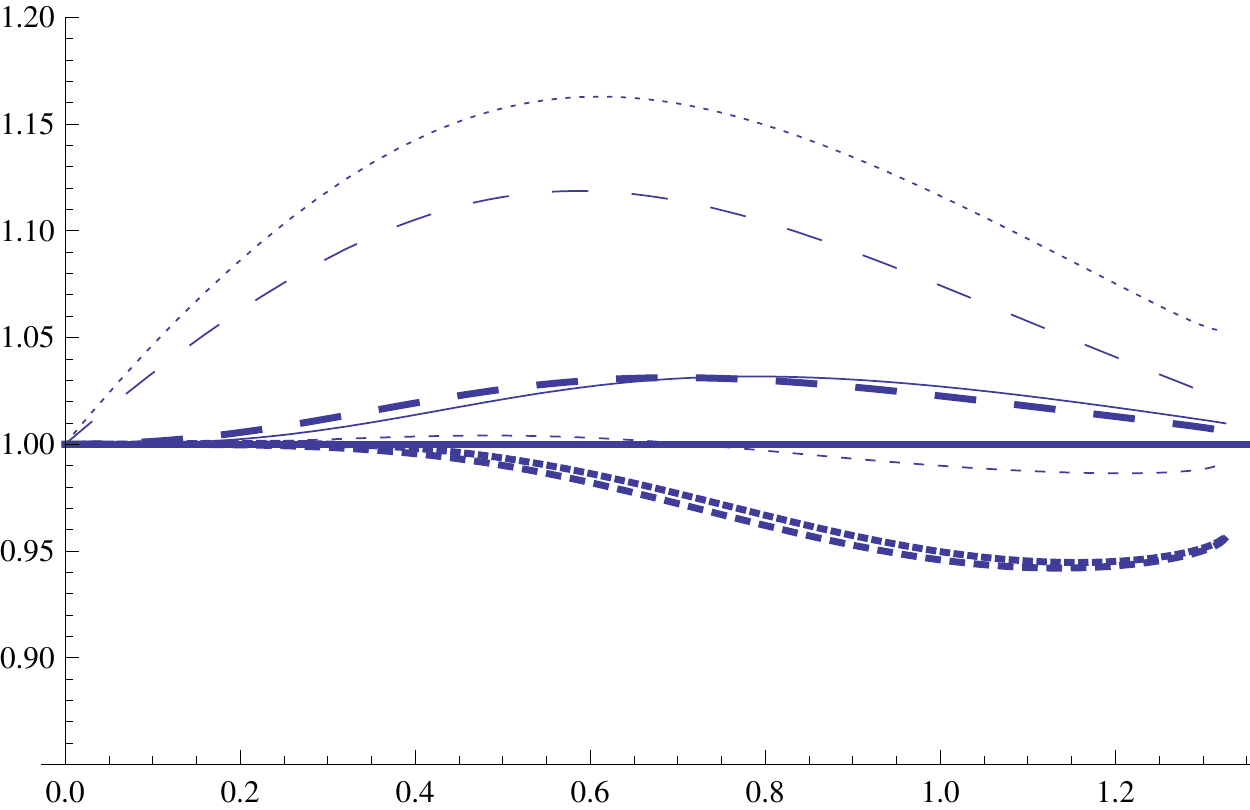} \epsfysize=5cm \epsfbox{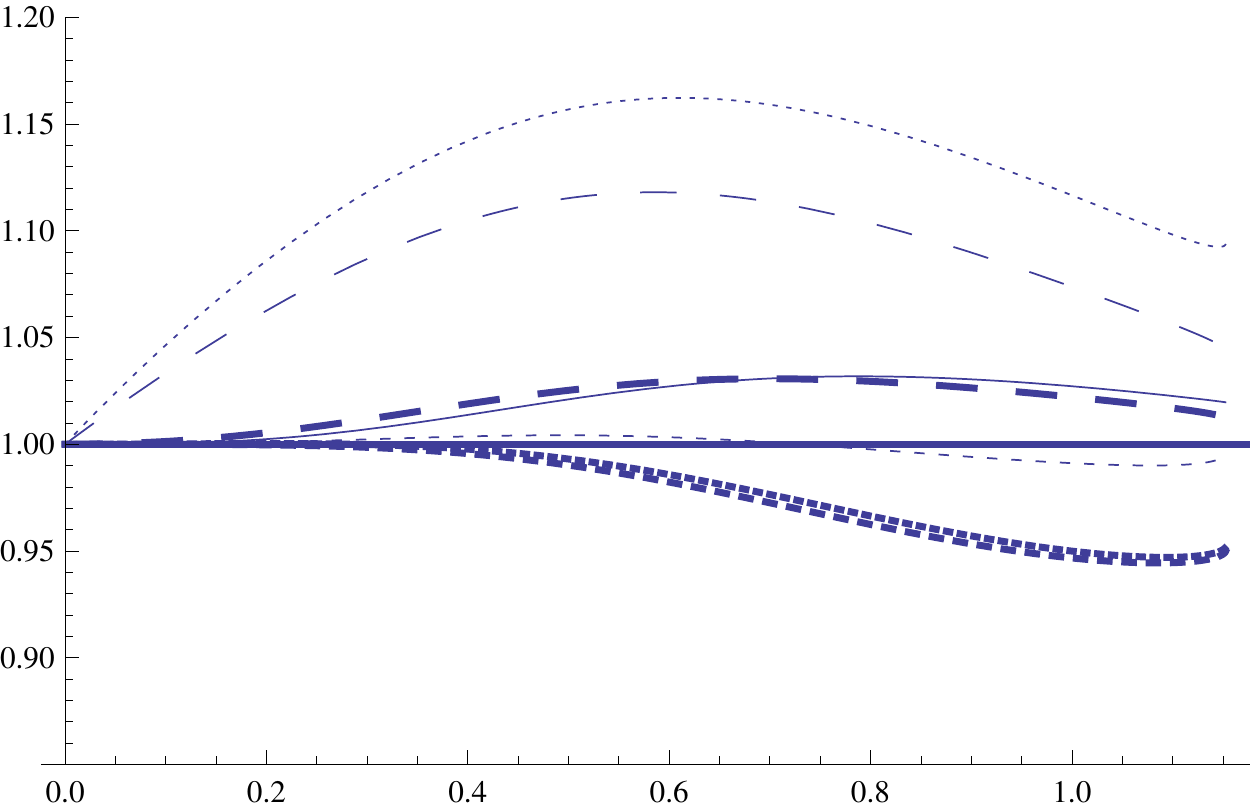}}
\end{center}
\caption{Graph of $F_{m,n}(g)/F_{3,1}(g)$ vs. $\tan^{-1}g$ for various $(m,n)$ in region I. 
The labels are as follows:  thin dots for $F_{0,0}$,  thick dots for $F_{1,0}$,
small thin dashes for $F_{2,0}$, small thick dashes for $F_{3,0}$, large thin dashes for
$F_{0,1}$, large thick dashes for $F_{1,1}$, continuous thin line for $F_{2,1}$ and
continuous thick line for $F_{3,1}$. The four graphs, clockwise from top left, correspond to
$r_H=1,1.25,1.5$ and 2 respectively.
\label{finter1}}
\end{figure}

In Fig.\ref{finter1} we have plotted the ratios of $F_{m,n}$ to $F_{3,1}$ as a function of
$g$ for four different values
of $r_H$  in region I.
As we can see, except for $F_{0,0}$, all other $F_{m,n}$'s remain within about 10\% of
$F_{3,1}$ over the entire allowed range of $g$ in region I. 
This suggests that $F_{3,1}$
gives the actual mass of the particle within about 10\% error over the entire range of
parameter space of region I (and hence also of region III). We shall return to a discussion of
regions II and IV in the next section.

\sectiono{Stability analysis} \label{sstability}

\begin{figure}
\vskip -1in
\begin{center}
\epsfysize=5cm
\hbox{\epsfbox{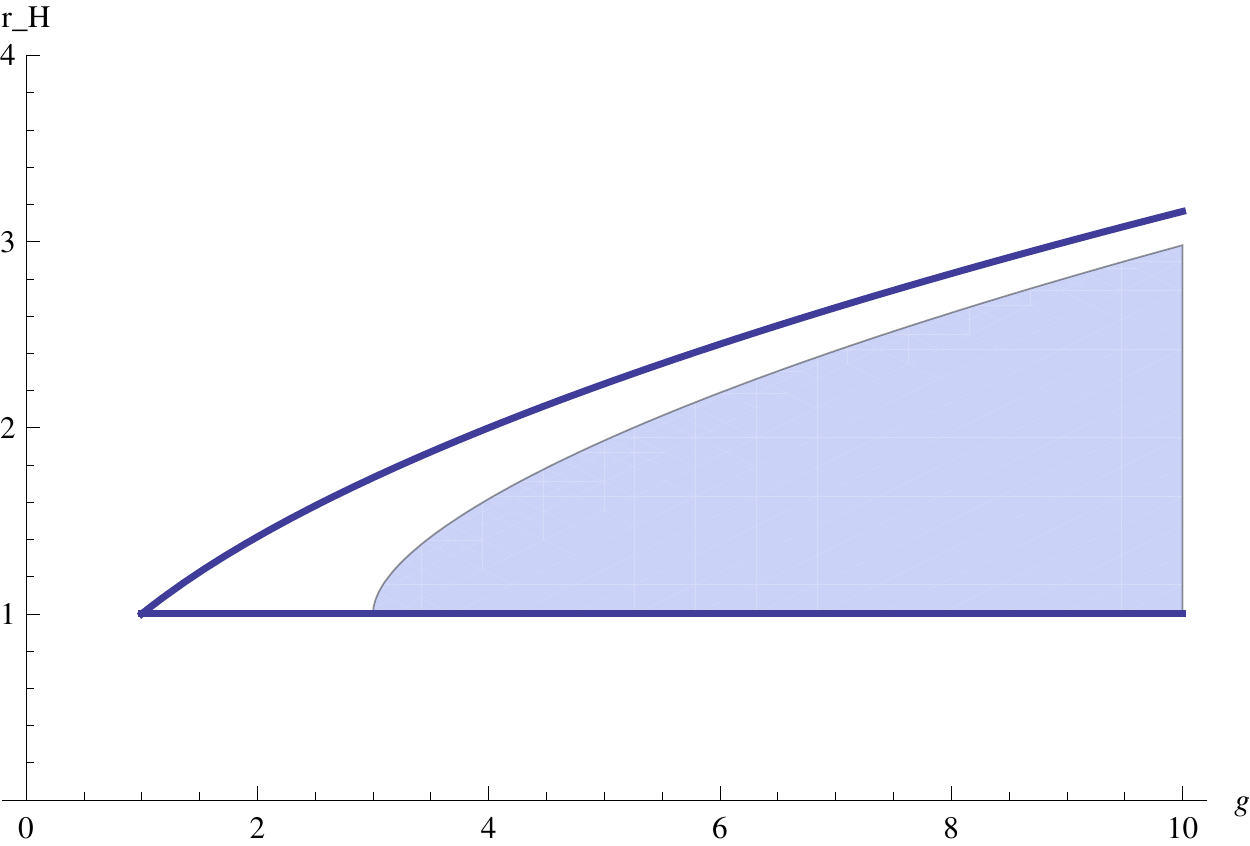} \epsfysize=5cm \epsfbox{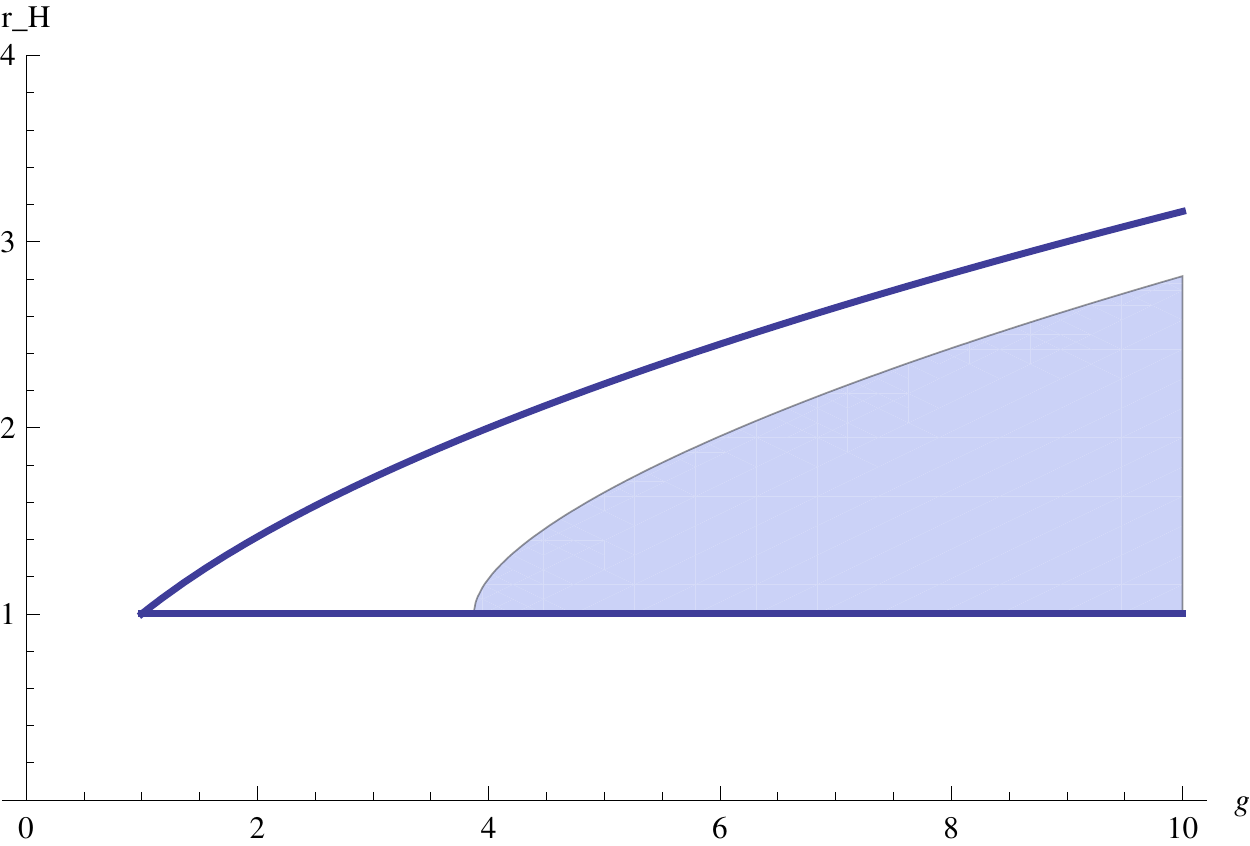}}
\epsfysize=5cm
\hbox{\epsfbox{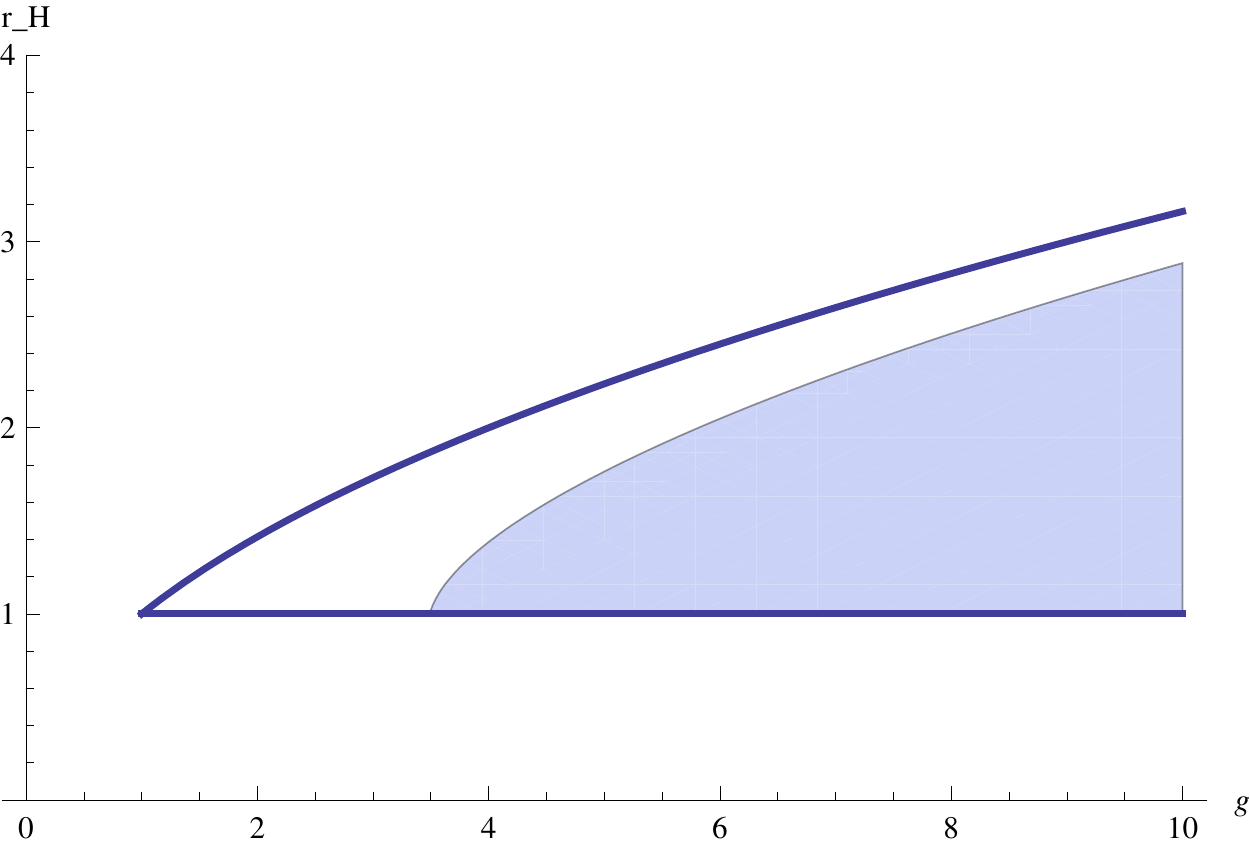} \epsfysize=5cm \epsfbox{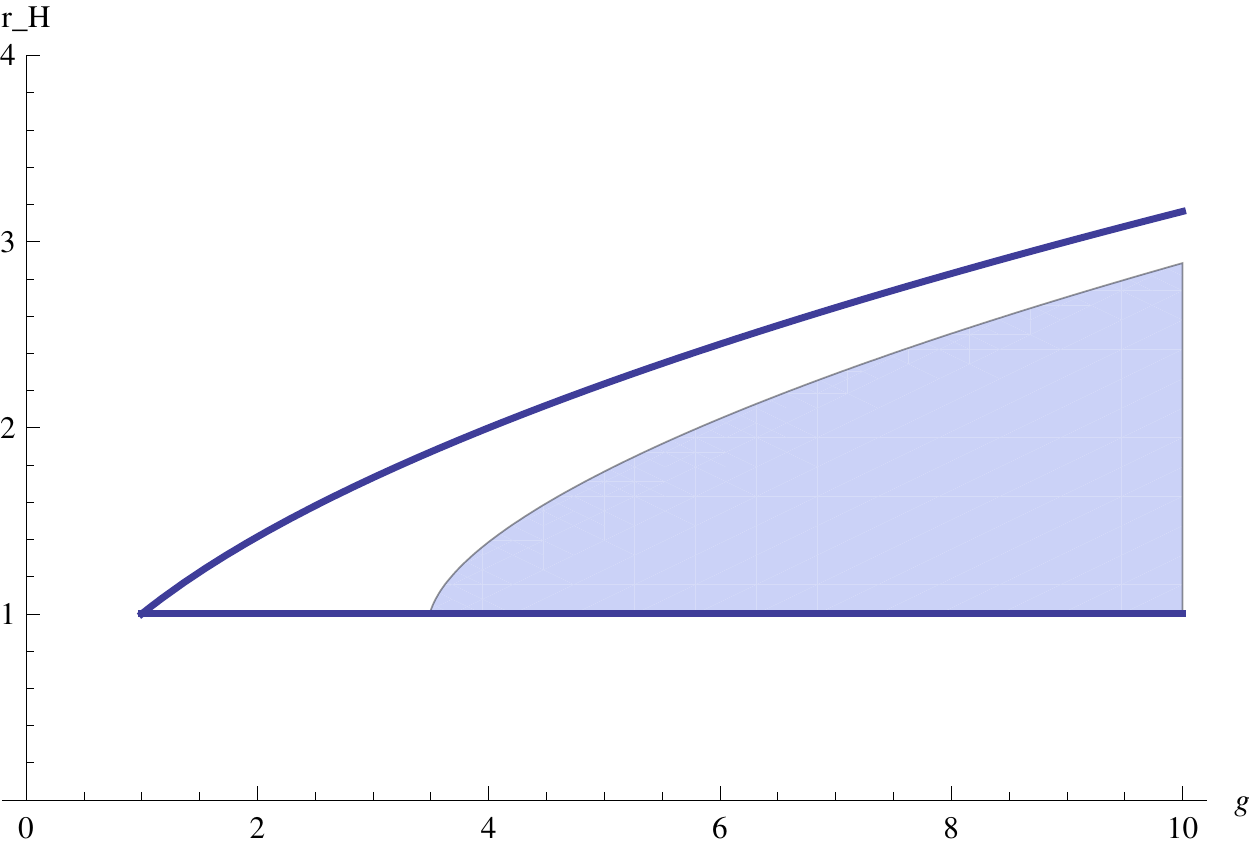}}
\end{center}
\caption{Region of instability of $F_{m,0}$ for $0\le m\le 3$ in the  region II of the $g-r_H$ plane.
Clockwise from top left the diagrams are based on the interpolating function $F_{0,0}$, $F_{1,0}$,
$F_{2,0}$ and $F_{3,0}$ respectively.
\label{f1}}
\end{figure}

Let $(n,w)$ denote the momentum and winding numbers of a heterotic string state
along the compact circle.
The non-BPS state carries the same quantum numbers as that of an heterotic string state
with quantum numbers $(n,w)=(1,1)$ in the spinor representation of SO(32), a state
with quantum number $(n,w) = (0,-1)$ in the adjoint/singlet representation of SO(32) and
a state with quantum number $(n,w)=(-1,0)$ in the singlet/adjoint representation.\footnote{We
could also consider decay into $(n,w)=(1,1)$ in the spinor representation and $(n,w)=(-1,-1)$
in the adjoint or singlet representation, or $(n,w)=(1,1)$ in the spinor representation,
$(n,w)=(1,-1)$ in the singlet representation and $(n,w)=(-2,0)$ in the singlet or adjoint
representation. In each of these cases the total mass of the decay products is  the
same as  that given on the right hand side of \refb{embps}.}
The total
mass of this state in the heterotic string metric for $r_H>1$ is
\be \label{embps}
M^H_{BPS} = \left( r_H + {1\over r_H}\right) + r_H + {1\over r_H}
= 2\left( r_H + {1\over r_H}\right)\, .
\ee
In the Einstein metric this is given by
\be 
M^E_{BPS} = (g_H)^{1/4} M^H_{BPS} = 2^{15/8} \pi^{7/8} g^{1/4}\, \left( r_H + {1\over r_H}\right)\, .
\ee
After taking into account the normalization \refb{edefF} we get
\be \label{efbps}
F_{BPS}(g) \equiv 2^{-15/8} \pi^{-7/8} M^E_{BPS} = g^{1/4} \, \left( r_H + {1\over r_H}\right)
= 2^{1/2} g^{3/4} \left( r_I + {1\over 2\, g\, r_I}\right)\, .
\ee
This expression is not renormalized. Furthermore it is manifestly invariant under heterotic T-duality
transformation \refb{etdual}.

\begin{figure}
\vskip -1in
\begin{center}
\epsfysize=5cm
\hbox{\epsfbox{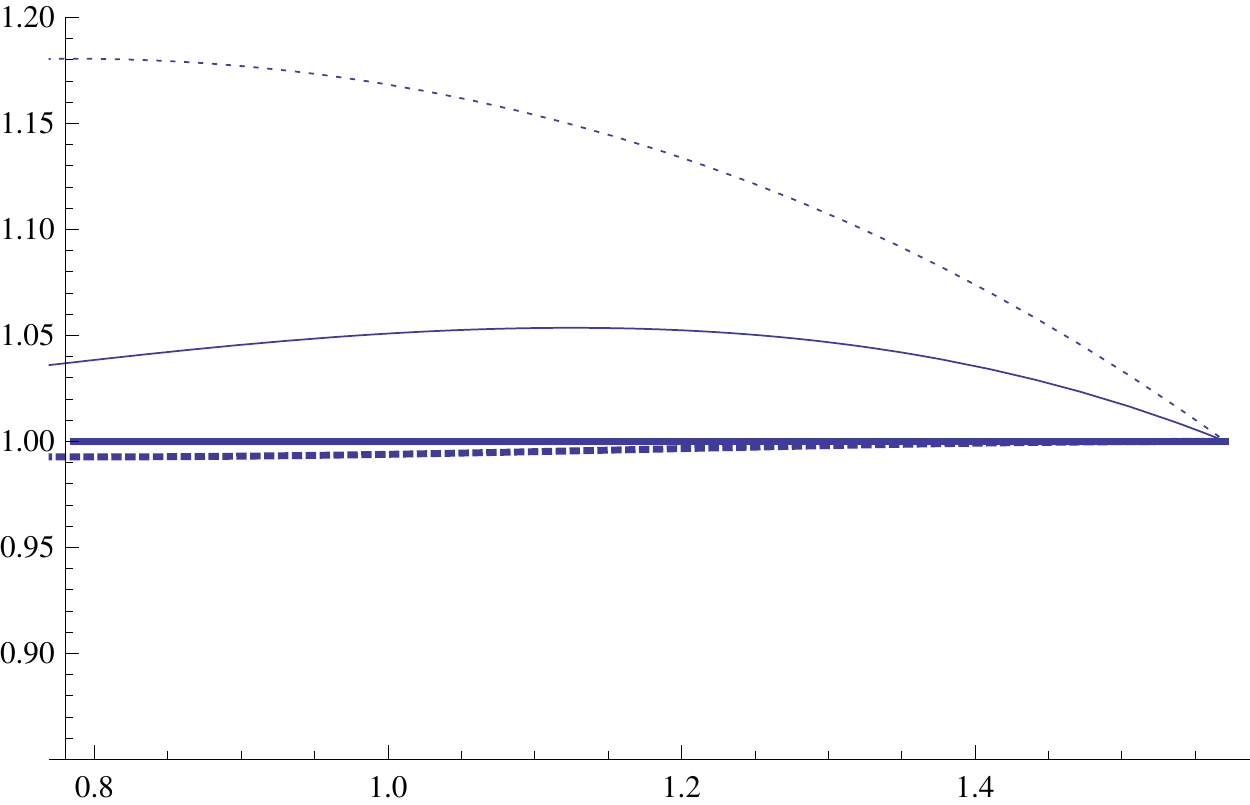} \epsfysize=5cm \epsfbox{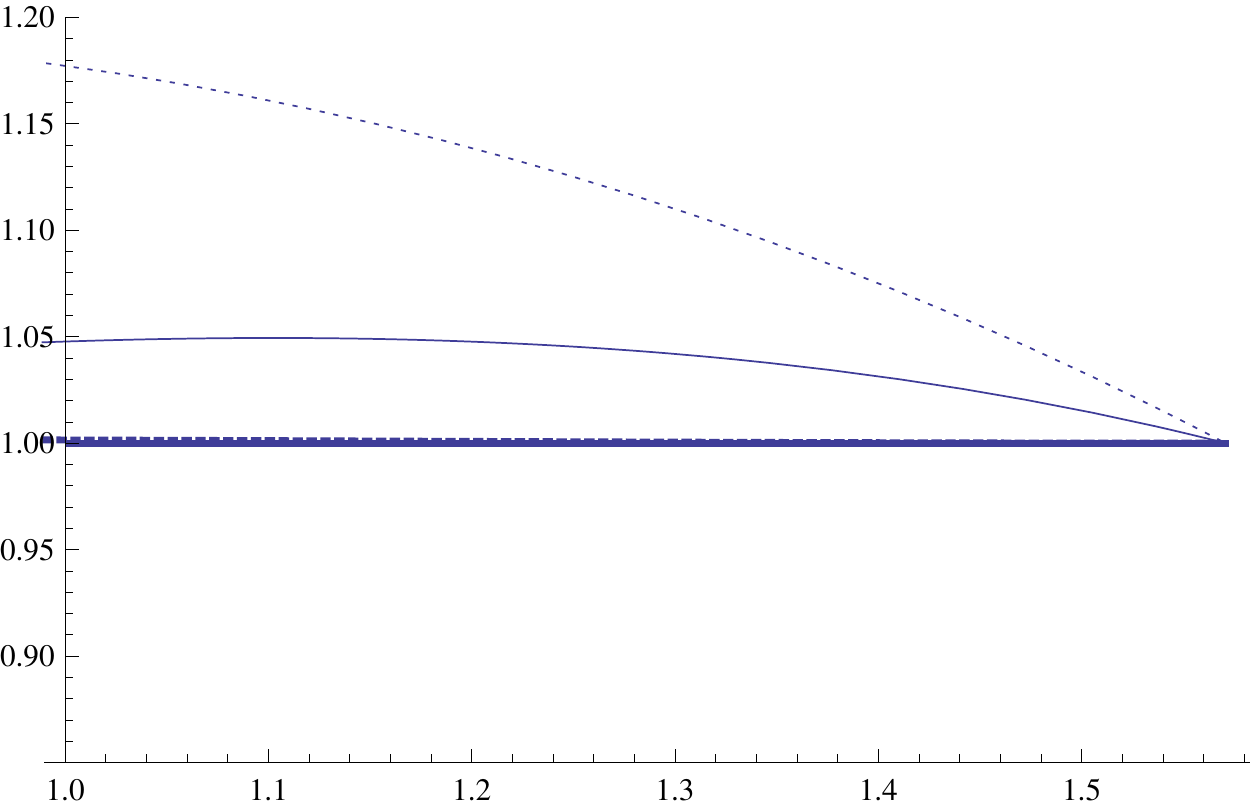}}
\epsfysize=5cm
\hbox{\epsfbox{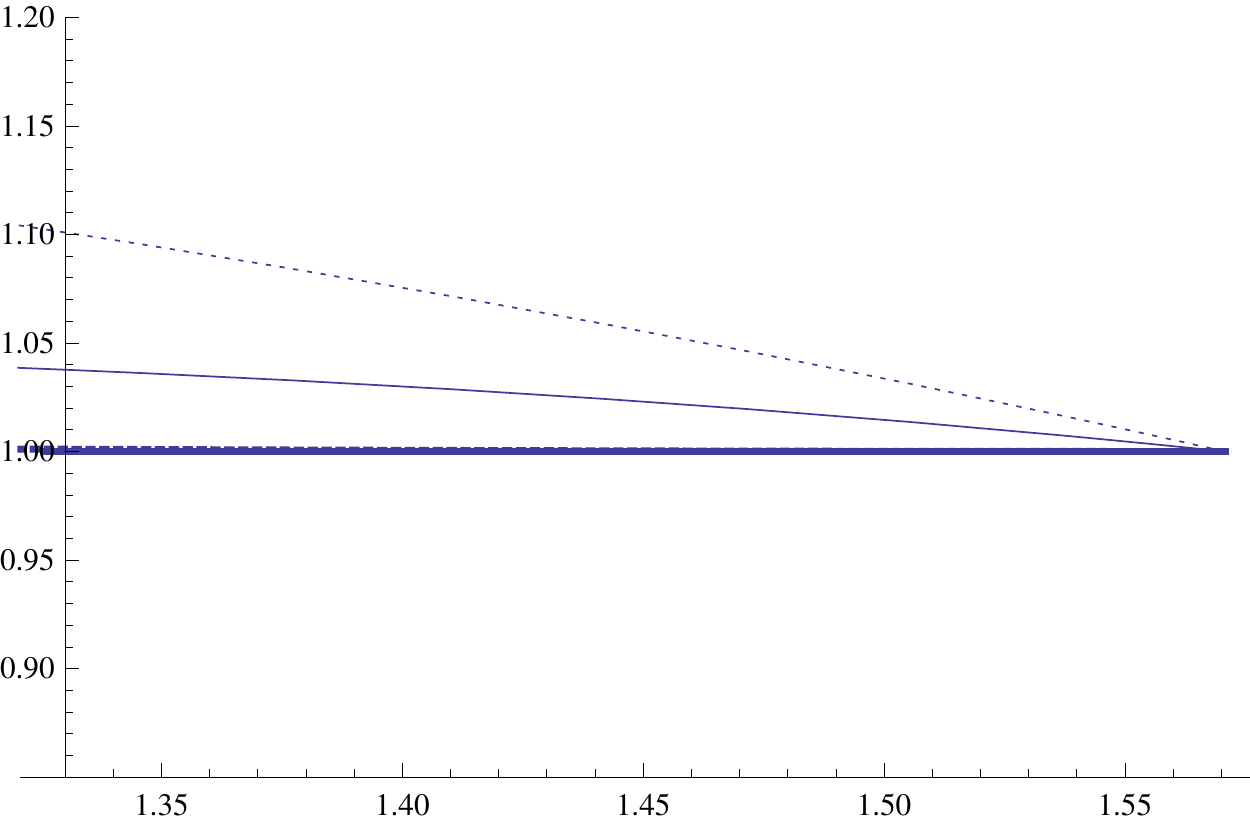} \epsfysize=5cm \epsfbox{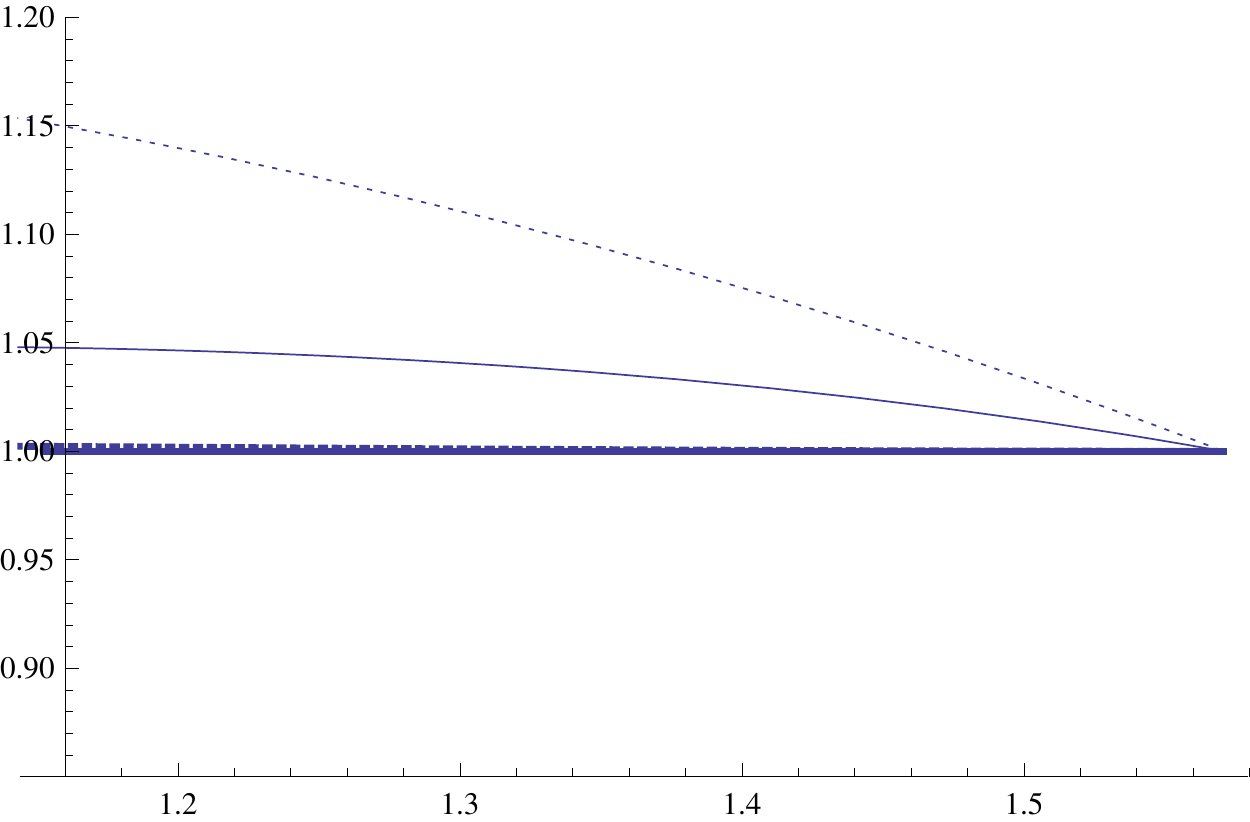}}
\end{center}
\caption{Graph of $F_{m,0}(g)/F_{3,0}(g)$ vs. $\tan^{-1}g$ for various $m$ in region II. 
The labels are as follows:  thin dots for $F_{0,0}$,  thick dots for $F_{1,0}$,
continuous thin line for $F_{2,0}$ and
continuous thick line for $F_{3,0}$. The four graphs, clockwise from top left, correspond to
$r_H=1,1.25,1.5$ and 2 respectively.
\label{finterII1}}
\end{figure}

In the $(m,n)$ approximation the non-BPS particle is stable when its mass is less than the total
mass of the BPS constituents to which it can decay. In regions I and II this requires
\be \label{estable}
F_{m,n}(g) < g^{1/4} \, \left( r_H + {1\over r_H}\right)\, .
\ee
In regions III and IV the left hand side is replaced by $\wt F_{m,n}(g)$, but the results in these
regions are related to those in regions I and II respectively by heterotic T-duality. Now one can
check explicitly that in region I all the $F_{m,n}(g)$'s satisfy \refb{estable}, showing that whatever
approximation we use, the non-BPS state is stable in this region. By heterotic T-duality the same
result holds in region III. In region II only $F_{m,0}$ approximations make sense. In Fig.~\ref{f1} we
have shown by the shaded region the region of instability of the non-BPS state in different
approximations. As we can see, these regions are not too different from each other, indicating that
this is a fairly good approximation to the true region of instabiity of the non-BPS state in region II.
The region of instability in region III can be found from this using heterotic T-duality transformation.

One point worth noticing is that in each of these plots, there is a narrow strip of region II where the
non-BPS state is stable. This may seem a bit surprising at first since in the whole of region II
perturbative open string theory describing the non-BPS D0-brane develops a tachyon. Note however
that this is true only in tree level open string theory which corresponds to $g\to\infty$ limit of this
diagram. Indeed in this limit the strip width reduces to zero showing that the D0-brane becomes
unstable as soon as we cross the upper boundary of region II. At finite $g$ however the tachyon
mass$^2$ itself may get corrected and hence the tachyon may not develop as soon as we cross
the upper boundary. We cannot do this analysis directly in type I string theory
since at present it is not understood how to
carry out open string perturbation theory in the presence of a tree level tachyon.
Instead we have checked the stability
by comparing the mass of the unstable brane with the total mass of the decay products, and
arrived at
Fig.~\ref{f1}.  

It is natural to ask whether one can reliably determine the mass of the non-BPS particle 
in the white part of region II where it is expected
to be stable. This can be done via the interpolating function $F_{m,0}$. 
Fig.\ref{finterII1} shows the ratios of $F_{m,0}$ to $F_{3,0}$ in region II.
As we can see from this graph, the ratios remain within about 10\% of unity except for
$F_{0,0}$, 
indicating that the interpolating
formul\ae\ based on $F_{3,0}$ 
determines the actual mass of the stable non-BPS particle even inside region II
to within about 10\% accuracy.

\sectiono{Compactification on higher dimensional tori} \label{shigher}

In this section we shall briefly discuss the generalization of the above analysis to
type I / SO(32) heterotic string theory compactified on $T^d$ -- a $d$ dimensional 
torus. We shall refrain from switching on any gauge field background so as to
keep SO(32) gauge group unbroken, but allow generic values of the other moduli. This
corresponds to choosing arbitrary constant metric $G_{Hmn}$ and NS-NS 2-form field
$B_{Hmn}$ along $T^d$ in heterotic description, and arbitrary constant metric
$G_{Imn}$ and RR 2-form field $C_{Imn}$ in the type I description. Generalization of
\refb{erirh} relating the two sets of moduli are
\be
G_{Hmn} = 2\, g\, G_{Imn}, \quad B_{Hmn} = C_{Imn}\, .
\ee
The weak and strong coupling expansions  take the same form as in \refb{expan} and
the
interpolation formula takes the form of \refb{einterpol} with the dependence on $r_H$ now
generalized to dependence on $G_{Hmn}$ and $B_{Hmn}$ and the dependence on $r_I$
generalized to dependence on $G_{Imn}$. The procedure for constructing the coefficients
$a_i$ and $b_i$ in \refb{einterpol} is the same as that for $S^1$ compactification.

The analog of the strong coupling expansion \refb{open}, \refb{efs1g} now takes the form:
\ben
F^S_1(g) &=& g^{\frac{3}{4}}\bigg(1+K_S\, g^{-1}\bigg),   \nonumber \\
K_S &\equiv& -2^{-5/2}
(8\pi^2)^{-\frac{1}{2}} \lim_{\Lambda \to \infty} \lim_{\epsilon\to 0} \Bigg[\int_{\eps}^{\Lambda}s^{-\frac{3}{2}}ds\Bigg\{\frac{1}{2}\Bigg(\sum_{\vec n} 
\tilde{q}^{2G_{Ik\ell} n^k n^\ell}\Bigg)\Bigg(\frac{f_3(\tilde{q})^8}{f_1(\tilde{q})^8}-\frac{f_2(\tilde{q})^8}{f_1(\tilde{q})^8}\Bigg) 
\cr 
 && +16\sqrt{2}\frac{f_2(\tilde{q})^9f_1(\tilde{q})}{f_4(\tilde{q})^9f_3(\tilde{q})}-16\sqrt{2}\frac{f_3(\tilde{q})^9f_1(\tilde{q})}{f_4(\tilde{q})^9f_2(\tilde{q})}\Bigg\} 
 \cr 
 && 
 +\int_{\eps/4}^{\Lambda}s^{-\frac{3}{2}}ds\Bigg\{ 2^{\frac{5}{2}}(1-i)\frac{f_3(i\tilde{q})^9f_1(i\tilde{q})}{f_2(i\tilde{q})^9f_4(i\tilde{q})} -2^{\frac{5}{2}}(1+i)\frac{f_4(i\tilde{q})^9f_1(i\tilde{q})}{f_2(i\tilde{q})^9f_3(i\tilde{q})}\Bigg\}\Bigg]\, ,
\een
where the sum over $\vec n$ refers to sum over $d$ integers $(n^1, \cdots n^d)$ labelling the
winding numbers of open strings along the $d$ circles. Similarly the weak coupling
result \refb{weak1} now takes the form
\ben\label{weakgen}
K_W=&&-\frac{1}{64\pi}\int d^2\tau \int d^2z \Bigg\{\sum_{\nu'}\{\overline{\vartheta_{\nu'}(\frac{z}{2})^{16}} \}(\overline{\eta(\tau)})^{-18}(\eta(\tau))^{-6}\Bigg(\frac{\vartheta_{11}(z)}{\overline{\vartheta_{11}(z)}}\Bigg)^2\Bigg\}\cr
&&\textrm{exp}\left(-\frac{4\pi z_2^2}{\tau_2}\right)
(\tau_2)^{(d-10)/2}(\det G_H)^{-1/2}\nonumber \\
&& \sum_{\vec n,\vec w}\textrm{exp}\bigg[-
\pi \tau_2 \Big\{ (G_H^{-1})_{k\ell} n^k n^\ell + (G_H - B_H G_H^{-1}B_H)_{k\ell} w^k w^\ell
+ 2 (G_H^{-1}B_H)_{k\ell} n^k w^\ell\Big\}\nonumber \\
&&  - 2\pi i \tau_1 n^k w^k
\bigg]\, ,
\een
where the sum over $\vec n$ and $\vec w$ respresent the sum over $d$ momentum quantum numbers
$(n^1,\cdots n^d)$ and  $d$ winding numbers $(w^1,\cdots w^d)$.

As in the case of $S^1$ compactification, we shall find that on the strong coupling side
the computation of $K_S$ suffers from tachyonic divergence when $2G_{Ik\ell} n^k n^\ell$ becomes
less than 1 for any non-zero $\vec n$. Inside this region we need to use only the zeroeth order
result on the strong coupling side. On the weak coupling side,
when the size of the torus $T^d$ is
small, $K_W$ computed from \refb{weakgen} becomes large signalling an apparent breakdown of
perturbation theory. The remedy is to use a T-duality transformation and use the T-dual 
variables. 
In fact here we have a large group  $O(6,6;\ZZZ)$ of T-duality transformation acting on the
moduli space. 
We need to identify the analog of the regions I and II in Fig.~\ref{f3} in which we carry out
the actual computation and
interpolation and then extend the result to the rest of the moduli space using heterotic T-duality
invarinace. 
Since in the compactified heterotic 
string theory the effective coupling constant is given by $g^2 / \sqrt{\det G_H}$, 
the natural analog of regions I and II will be to pick that domain in the moduli space for which 
$\sqrt{\det G_H}$ takes the maximum possible value -- \i.e.\ given any point inside such a 
domain, any of its T-dual image should have lower value of $\sqrt{\det G_H}$. Once we have identified
such a domain we then divide this into the two regions I and II depending on whether
$2G_{Ik\ell} n^k n^\ell$ lies above 1 for all $\vec n$ or not. The rest of the analysis would then
proceed as in the case of $S^1$ compactification.

\sectiono{Conclusion} \label{sconc}

In this paper we have analyzed the mass formula for stable non-BPS state in type I /
SO(32) heterotic string theory compactified on a circle using the interpolation formula between
the strong and weak coupling results. Our analysis indicates that the interpolation formula
determines the mass of the state within 10\% accuracy over the entire moduli space. We also
determine the region of stability of the particle based on the mass formula and discuss generalization
of the analysis for generic toroidal compactification.

In recent times there has been significant developments in resumming 
perturbation theory\cite{0207349,0405279,1106.5922,1206.1890,
1210.2423,1210.3646,1302.5138,1308.0127,1308.1108,1308.1115,1308.1695,1309.0437}. 
It will be interesting to see if the interpolation between strong
and weak coupling results can be combined efficiently with these resummation techniques
to get a better understanding of physical quantities at intermediate values of coupling.

\bigskip

{\bf Acknowledgement:}  We would like to thank  Rajesh Gopakumar and Dileep Jatkar
for useful discussions.
This work was
supported in part by the 
DAE project 12-R\&D-HRI-5.02-0303. The work of A.S. was also supported in
part by the
J. C. Bose fellowship of 
the Department of Science and Technology, India.


\small

\baselineskip 10pt

\begin{thebibliography}{99}

\bibitem{1209.5461} 
  E.~Witten,
  ``Superstring Perturbation Theory Revisited,''
  arXiv:1209.5461 [hep-th].



\bibitem{1304.0458} 
  A.~Sen,
  ``S-duality Improved Superstring Perturbation Theory,''
  arXiv:1304.0458 [hep-th].

\bibitem{1306.3228}
  C.~Beem, L.~Rastelli, A.~Sen and B.~C.~van Rees,
  ``Resummation and S-duality in N=4 SYM,''
  arXiv:1306.3228 [hep-th].

\bibitem{1307.3689}
  T.~Banks and T.~J.~Torres,
  ``Two Point Pade Approximants and Duality,''
  arXiv:1307.3689 [hep-th].
  
  \bibitem{1310.3757}
  L.~F.~Alday and A.~Bissi, 
``The superconformal bootstrap
for structure constants,"  
arXiv:1310.3757 [hep-th].

\bibitem{0706.1555}
V.~Asnin, D.~Gorbonos, S.~Hadar, B.~Kol, M.~Levi and U.~Miyamoto,
  ``High and Low Dimensions in The Black Hole Negative Mode,''
  Class.\ Quant.\ Grav.\  {\bf 24}, 5527 (2007)
  [arXiv:0706.1555 [hep-th]].

\bibitem{kleinert} 
  H.~Kleinert and V.~Schulte-Frohlinde,
  ``Critical properties of  $\phi^4$-theories,''
  River Edge, USA: World Scientific (2001) 489 p


\bibitem{9510017}
  J.~Polchinski,
  Phys.\ Rev.\ Lett.\  {\bf 75} (1995) 4724
  [hep-th/9510017].

\bibitem{9903123} 
  M.~Frau, L.~Gallot, A.~Lerda and P.~Strigazzi,
  ``Stable nonBPS D-branes in type I string theory,''
  Nucl.\ Phys.\ B {\bf 564}, 60 (2000)
  [hep-th/9903123].

\bibitem{0003022}
M. Frau, L. Gallot, A. Lerda and P. Strigazzi,
``Stable non-BPS D branes of type I,''
hep-th/0003022.

\bibitem{0012167}
M. Frau, L. Gallot, A. Lerda and P. Strigazzi,
``D-branes in type I string theory,''
  Fortsch.\ Phys.\  {\bf 49}, 503 (2001)
  [hep-th/0012167].

\bibitem{9904207} 
  A.~Sen,
  ``NonBPS states and Branes in string theory,''
  hep-th/9904207.

\bibitem{0207349} 
  M.~Stingl,
  ``Field theory amplitudes as resurgent functions,''
  hep-ph/0207349.

\bibitem{0405279} 
  U.~D.~Jentschura and J.~Zinn-Justin,
  ``Instantons in quantum mechanics and resurgent expansions,''
  Phys.\ Lett.\ B {\bf 596}, 138 (2004)
  [hep-ph/0405279].


\bibitem{1106.5922} 
  I.~sAniceto, R.~Schiappa and M.~Vonk,
  ``The Resurgence of Instantons in String Theory,''
  Commun.\ Num.\ Theor.\ Phys.\  {\bf 6}, 339 (2012)
  [arXiv:1106.5922 [hep-th]].


\bibitem{1206.1890} 
  P.~C.~Argyres and M.~Unsal,
  ``The semi-classical expansion and resurgence in gauge theories: new perturbative, instanton, bion, and renormalon effects,''
  JHEP {\bf 1208}, 063 (2012)
  [arXiv:1206.1890 [hep-th]].


\bibitem{1210.2423} 
  G.~V.~Dunne and M.~Unsal,
  ``Resurgence and Trans-series in Quantum Field Theory: The CP(N-1) Model,''
  JHEP {\bf 1211}, 170 (2012)
  [arXiv:1210.2423 [hep-th]].


\bibitem{1210.3646} 
  G.~V.~Dunne and M.~Unsal,
  ``Continuity and Resurgence: towards a continuum definition of the CP(N-1) model,''
  Phys.\ Rev.\ D {\bf 87}, 025015 (2013)
  [arXiv:1210.3646 [hep-th]].


\bibitem{1302.5138} 
  R.~Schiappa and R.~Vaz,
  ``The Resurgence of Instantons: Multi-Cuts Stokes Phases and the Painleve II Equation,''
  arXiv:1302.5138 [hep-th].

\bibitem{1308.0127} 
  A.~Cherman, D.~Dorigoni, G.~V.~Dunne and M.~Unsal,
  ``Resurgence in QFT: Unitons, Fractons and Renormalons in the Principal Chiral Model,''
  arXiv:1308.0127 [hep-th].


\bibitem{1308.1108} 
  G.~Basar, G.~V.~Dunne and M.~Unsal,
  ``Resurgence theory, ghost-instantons, and analytic continuation of path integrals,''
  arXiv:1308.1108 [hep-th].


\bibitem{1308.1115} 
  I.~sAniceto and R.~Schiappa,
  ``Nonperturbative Ambiguities and the Reality of Resurgent Transseries,''
  arXiv:1308.1115 [hep-th].


\bibitem{1308.1695} 
  R.~C.~Santamar'a, J.~Ž D.~Edelstein, R.~Schiappa and M.~Vonk,
  ``Resurgent Transseries and the Holomorphic Anomaly,''
  arXiv:1308.1695 [hep-th].


\bibitem{1309.0437} 
  M.~Garay, A.~de Goursac and D.~van Straten,
  ``Resurgent Deformation Quantisation,''
  arXiv:1309.0437 [math-ph].


\end{thebibliography}
\end{document}